\documentclass[12pt]{article}

\usepackage{amsmath,amssymb,amsthm,color,graphicx}
\usepackage{mathpazo}
\usepackage{mathtools}
\usepackage{xcolor}

\usepackage{comment}

\usepackage{euscript}
\usepackage[top=1.25in,bottom=1.25in,left=1.25in,right=1.25in]{geometry}

\usepackage{enumerate}
\usepackage[linewidth=1pt]{mdframed}

\usepackage{textcomp}
\usepackage{microtype}
\usepackage{fullpage}
\usepackage{xspace}
\usepackage{tikz}  % Used for the extended footnotes if needed
\usepackage{floatflt}
\usepackage{wrapfig}
\usepackage{hyperref}
\usepackage{cleveref}

\usepackage{booktabs}

\usepackage{subcaption}

\usepackage{scalerel}

\usepackage[backend=bibtex,maxbibnames=99,sortcites=true]{biblatex}
\renewbibmacro*{doi+eprint+url}{
    \printfield{doi}
    \newunit\newblock
    \iftoggle{bbx:eprint}{
        \usebibmacro{eprint}
    }{}
    \newunit\newblock
    \iffieldundef{doi}{
        \usebibmacro{url+urldate}}
        {}
    }
\hypersetup{
    colorlinks,
    linkcolor={red!50!black},
    citecolor={blue!50!black},
    urlcolor={blue!80!black}
}
\newcommand{\figlab}[1]{\label{fig:#1}}

\newcommand{\figref}[1]{Figure~\ref{fig:#1}}

\newcommand{\opt}{\textrm{OPT}}
\newcommand{\etal}{et al.}

\newcommand{\newinf}{\mathop{\mathrm{inf}\vphantom{\mathrm{sup}}}}

\newtheorem{theorem}{Theorem}[section]

\newtheorem{observation}{Observation}[theorem]

\graphicspath{{./figs/}}

\newcommand*{\reals}{{\mathbb R}}

\newcommand*{\Reals}{\reals}

\makeatletter
\DeclareFontEncoding{LS1}{}{\noaccents@} 
\makeatother
\DeclareFontSubstitution{LS1}{stix}{m}{n}

\DeclareSymbolFont{symbolsSTIX}{LS1}{stixscr}{m}{n}
\DeclareSymbolFont{symbols2STIX}{LS1}{stixfrak}{m}{n}
\DeclareMathSymbol{\boxbox}{\mathbin}{symbols2STIX}{"B7}
\DeclareMathSymbol{\boxwhite}{\mathord}{symbolsSTIX}{"B8}
\DeclareMathSymbol{\boxblackbox}{\mathord}{symbolsSTIX}{"BA}

\newcommand{\ignore}[1]{}

\makeatletter
\long\def\@makecaption#1#2{
   \vskip 10pt
   \setbox\@tempboxa\hbox{{\footnotesize \textbf{#1.} #2}}
   \ifdim \wd\@tempboxa >\hsize         % IF longer than one line:
       {\footnotesize \textbf{#1.} #2\par}% THEN set as ordinary paragraph.
     \else                              %   ELSE  center.
       \hbox to\hsize{\hfil\box\@tempboxa\hfil}
   \fi}
\makeatother

\makeatletter
\DeclareRobustCommand\onedot{\futurelet\@let@token\@onedot}
\def\@onedot{\ifx\@let@token.\else.\null\fi\xspace}

\makeatother

\newcounter{problemNumber}
\def\O{{\cal O}}

\newcommand{\TwoDiscsMinSum}{\textbf{2Discs\-Min\-Sum}\xspace}

\newcommand{\TwoDiscsNoObstacles}{\textbf{2DiscsNoObstacles}\xspace}
\newcommand{\TwoSqrsMinSum}{\textbf{2SqrsMinSum}\xspace}
\newcommand{\ThreeDCaging}{\textbf{3DCaging}\xspace}
\newcommand{\TwoDCagingTrans}{\textbf{2DCagingTrans}\xspace}

\newcommand{\TwoDHoleInMinkowskiSum}{\textbf{2DHoleInMinkowskiSum}\xspace}

\newcommand{\ThreeInterlockeddFewerThanThirty}{\textbf{3DIntrelockedFewerThan30}\xspace}

\newcommand{\MDMTwoDPartitionFiniteTrans}
{\textbf{MDM2DPartition}\xspace}

\newcommand{\MDMTwoDAssemblyFiniteTrans}
{\textbf{MDM2DAssembly}\xspace}

\newcommand{\MDMTwoDReconfigurationUnitDiscsMove}
{\textbf{MDM2DReconfigurationUnitDiscsMove}\xspace}

\newcommand{\MDMTwoDReconfigurationUnitDiscsShrink}
{\textbf{MDM2DReconfigurationUnitDiscsShrink}\xspace}

\newcommand{\MDMTwoDReconfigurationUnitDiscsMinNumShrink}
{\textbf{MDM2DReconfigurationUnitDiscsMinNumShrink}\xspace}

\newcommand{\MDMCastingWithOnePartMold}
{\textbf{MDMCasting}\xspace}

\newcommand{\PartitioningThreeD}
{\textbf{PartitioningPolyhedra}\xspace}

\newcommand{\PartitioningPacking}
{\textbf{PartitioningForPacking}\xspace}

\newcommand{\UnlabeledMultiDiscInterpolate}
{\textbf{UnlabeledMultiDiscInterpolate}\xspace}

\newcommand{\UnlabeledMonotoneReconfiguration}
{\textbf{UnlabeledMonotoneReconfiguration}\xspace}

\newcommand{\ConvexCover}
{\textbf{ApproximateConvexCover}\xspace}

\newcommand{\ThreeHandedPartitionInfiniteTrans}
{\textbf{3HandedPartitionInfiniteTrans}\xspace}

\newcommand{\TwoDiscsMinCombinedMeasure}{\textbf{2Discs\-Min\-Combined\-Measure}\xspace}

\newcommand{\CompactTetrahedralization}{\textbf{CompactTetrahedralization}\xspace}

\newcommand{\BoundingVolumeHierarchy}{\textbf{BoundingVolumeHierarchy}\xspace}

\title{\bf Ten Problems in Geobotics\thanks{Work by M.A.~supported by Starting Grant 1054-00032B from the Independent Research Fund Denmark under the Sapere Aude research career programme and VILLUM Foundation grant 16582. Work by D.H.\  has been supported in part by the Israel Science
Foundation (grant no.~2261/23), 
by NSF/US-Israel-BSF (grant no.~2019754),
by the Blavatnik Computer Science Research Fund, 
and by the Shlomo Shmelzer Institute for 
Smart Transportation at Tel Aviv University.}}
\date{August 2024}
\author{Mikkel Abrahamsen \and Dan Halperin}

\begin{document}
\maketitle

\begin{abstract}
Robots sense, move and act in the physical world. It is therefore natural
that algorithmic problems in robotics and automation have a geometric component, often central to the problem. Below we
review ten challenging problems at the intersection of robotics and
computational geometry---let's call this intersection
\emph{Geobotics}. What is common to most of these problems is that the prevalent algorithmic techniques used in robotics do not seem suitable for solving them, or at least do not suggest quality guarantees for the solution.
Solving some of them, even partially, can shed light on
less well-understood aspects of computation in robotics.
\end{abstract}

\setcounter{section}{-1}

\section{Introduction}

Robotics has persistently raised interesting problems for computational geometry. A pioneering and exemplary case in point is the 1980s series of papers ``On the Piano Movers Problem'' by Schwartz and Sharir~\cite{schwartz1983pianoI,schwartz1983pianoII}, which laid out the theoretical foundation for the study of motion planning in robotics as well as many other applications. These were followed by a sequence of results studying special instances of motion-planning problems raising fundamental questions on the structure and complexity of arrangements of curves and surfaces (see, e.g., ~\cite{DBLP:journals/tcs/EdelsbrunnerGPPSS92,DBLP:journals/dcg/HalperinS95,hs-a-18,DBLP:journals/dcg/KedemLPS86}), or special types of Voronoi diagrams~\cite{DBLP:journals/dcg/LevenS87,DBLP:conf/stoc/ODunlaingSY83}.

A large set of results in automation have followed, again using and promoting the development of computational geometry techniques, for example in part orienting and fixturing  (see, e.g., ~\cite{DBLP:journals/comgeo/BerrettyGOS98,DBLP:conf/compgeom/RaoKG95,DBLP:conf/dagstuhl/StappenBGO00,DBLP:journals/algorithmica/StappenG00}) or in grasping  (see, e.g., ~\cite{DBLP:conf/isaac/CheongHS03,DBLP:journals/dcg/KirkpatrickMY92,DBLP:journals/ijrr/MarkenscoffNP90,DBLP:journals/comgeo/RaoG96}).

While this type of research moderately continued, the area of algorithms in robotics has undergone a major change in the mid 1990s, with the introduction of sampling based (SB, for short) techniques, most notably Probabilistic Road-Maps (PRM)~\cite{DBLP:journals/trob/KavrakiSLO96}, ensued by Rapidly exploring Random Trees (RRT)~\cite{DBLP:journals/ijrr/LaValleK01}, and their many variants~\cite{DBLP:journals/arcras/OrtheyCK24}. There are a bounty of methods in the algorithmic toolbox for robotics. However, we mention SB techniques in particular since many of the problems that we review below relate to motion of robots (or other objects), and this is an area that has been predominated by SB techniques for about three decades now. 

Sampling-based methods revolutionized the algorithmic approach in robotics. Many problems, which were previously practically insolvable, now became not only efficiently solvable, but also with only moderate programming efforts, assuming you have at your disposal good auxiliary tools for collision detection~\cite{CDChpaterLinManocha2018} and nearest-neighbor search (see, e.g., \cite{DBLP:conf/wafr/IchnowskiA18,DBLP:conf/visapp/MujaL09}). However, when the setting is tight, as it is for example in assembly planning, or packing, these methods often perform poorly, as their performance heavily relies on the denseness of the setting. They can run for a very long time without finding a solution, and it is not clear whether a solution does not exist or more time is needed. This issue becomes more pronounced when dealing with multi-robot or multi-body systems, and when striving for high quality, optimized, solutions. 

Machines learning (ML) techniques are becoming ubiquitous in robotics in recent years. Besides a conference, CoRL, dedicated to learning in robotics, these techniques preponderate the more traditional long-running conferences in the field; indeed in ICRA 2024, the two winners of the best paper award are focused on ML and data centered robotics. The open problems that we suggest here have not yet been successfully addressed with ML techniques, to the best of our knowledge. Therefore, we pose an overarching problem of whether any of the algorithmic questions reviewed below can be efficiently solved using ML.

For each problem, we provide the basic background including terminology and references in the respective sections. We do assume some familiarity with the basic terminology of motion planning, and in particular (the number of) the degrees of freedom of a system, and the notion of configuration space. For the fundamentals and terminology, consider the classical books by Latombe~\cite{DBLP:books/daglib/0068760} or LaValle~\cite{lavalle2006planning}, or the surveys in the Handbook on Discrete and Computational Geometry (3rd edition)~\cite{hks-r-18,hss-amp-18}.

Several problems in this review concern \emph{assembly planning}. For the needed algorithmic basics, consider, for example,
\cite[Section~51.2]{hks-r-18} or
the first three sections of the survey~\cite{DBLP:journals/algorithmica/HalperinLW00}.

\section{Shortest coordinated paths for two robots}\label{sec:shortest-coordinated}

Let $A$ and $B$ be two unit-disc robots (``roombas'') moving among polygonal obstacles $\O$ in the plane. We specify the \emph{configuration} (see Table~\ref{table:definitions-motion-planning}) of each robot by the coordinates of its center: When robot $R$ is placed with its center at the point $p\in \Reals^2$, we let $R(p)$ denote $\{x\in \Reals^2|  \lVert x-p \rVert <1\}$. Our goal is to plan a collision-free motion for the two robots from given free start configurations marked $A_0$ and $B_0$, to given free target configurations $A_1$ and $B_1$. Throughout the motion the robots should not collide with the obstacles, nor with one another. More formally, a solution to the coordinated motion for $A$ and $B$
is a plan $(\pi_A,\pi_B)$, where for each $R\in\{A,B\}$, we have $\pi_R:[0,1]\rightarrow \Reals^2$ with $\pi_R(0)=R_0, \pi_R(1)=R_1$, and for all $t\in[0,1]: R(\pi_R(t))\cap \O=\emptyset$
and 
$A(\pi_A(t))\cap B(\pi_B(t))=\emptyset$.
Notice that the obstacles $\O$ constitute a closed set and each robot is an open set, so in a valid motion plan the robots may touch, but never overlap each other or the obstacles.
It is not known how to efficiently find a coordinated motion plan that minimizes the total length of the paths, or if the problem is provably hard.
We formulate this as the following question.

\begin{table}  
\begin{center}
\begin{tabular}{l  p{.67\textwidth}}
\toprule
\textbf{Workspace}    & A subset $W$ of 2D or 3D physical space: $W \subset %\Bbb{R}^k$, $k = 2$ or $3$. \\
\mathbb{R}^k$, $k = 2$ or $3$. \\
\addlinespace
\textbf{Configuration}    & Any
  mathematical specification of the position and orientation of every rigid
  body composing a robot, relative to a fixed coordinate system.  The
  configuration of a single rigid body is also called a
  \textbf{placement} or a \textbf{pose}. \\
\addlinespace
\textbf{Configuration space}    &  The set of all configurations of
  a robot. For almost any robot, this set is a smooth manifold.  Given a robot $R$, we will let
  $R(q)$ denote the subset of the workspace occupied by $R$ at
  configuration $q$.\\
\addlinespace

\textbf{$\#$ of degrees of freedom}    & The dimension of the configuration space. \\
\bottomrule
\end{tabular}
\end{center}
\caption{Basic definitions in \textbf{robot} motion planning, adapted from~\cite{hks-r-18}.\label{table:definitions-motion-planning}}
\end{table}

\begin{quote}
	\textbf{Problem~\TwoDiscsMinSum}: Given two unit discs moving in the plane among polygonal obstacles with a total of $n$ vertices, devise an efficient algorithm to find a coordinated motion plan $(\pi^*_A,\pi^*_B)$ for the discs from free start to free target configurations such that the total length of the paths of the two robots is the smallest among all possible solutions, $$(\pi^*_A,\pi^*_B)={\rm argmin}_{(\pi_A,\pi_B){\rm {\text -}plan}} |\pi_A|+|\pi_B|,$$ if such a motion exists.
 Alternatively, show that the problem is NP-hard.
\end{quote}

If we relax the requirement that the paths are optimal (MinSum) in \textbf{Problem~\TwoDiscsMinSum}, and just seek to produce feasible solution paths (where the robots do not collide with the obstacles nor with one another along the way), then an efficient solution due to Sharir and Sifrony has been known for a long time~\cite{DBLP:journals/amai/SharirS91}, running in $O(n^2)$ time.

Interestingly, if we keep the optimality requirement (MinSum) and drop the obstacles, namely the two disc robots move freely in the plane and just have to avoid collision with each other, the problem is still non-trivial and was  open 
for a long time until settled several years back by Kirkpatrick and Liu~\cite{DBLP:conf/cccg/KirkpatrickL16,DBLP:journals/corr/KirkpatrickL16}. They analyze the shape of the optimal coordinated paths, and show that these paths consist of at most six line segments and circular arcs. See \figref{LiuKirkpatrickExample} for an illustration.

\begin{figure}
\centering
\includegraphics[page=2]{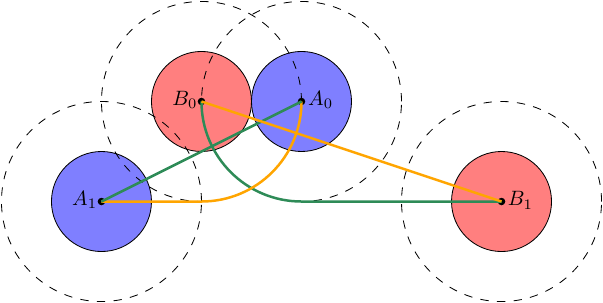}
\caption{Two optimal solutions of an instance of \textbf{Problem~\TwoDiscsMinSum} in the absence of obstacles, one in orange and one in green. The shaded discs are the unit-disc robots at their start and target configurations, and the dotted circles are of radius $2$ around the start and target configurations.
Example following~\cite{DBLP:journals/corr/KirkpatrickL16}.}
\figlab{LiuKirkpatrickExample}
\end{figure}

The solution for the MinSum no-obstacle setting has recently been adapted to the case of two squares by Esteban, Halperin, Ru{\'{\i}}z, Sacrist{\'{a}}n and Silveira~\cite{DBLP:conf/wads/EstebanHRSS23}.
The solution for two squares is more involved than for two discs since some symmetries (relative orientation of the two robots) that can be exploited in the latter do not exist in the former.

Back to the full-blown \textbf{Problem~\TwoDiscsMinSum} seeking an optimal solution in the presence of obstacles, we call its two-square variant, where instead of a unit disc each robot is a translating-only axis-parallel unit square,  \textbf{Problem~\TwoSqrsMinSum}.
There is no known polynomial-time algorithm for \textbf{Problem~\TwoSqrsMinSum}, and here as well we do not know that the problem is provably hard. Recently, Agarwal, Halperin, Sharir and Steiger~\cite{DBLP:conf/soda/AgarwalHSS24} have presented a PTAS for this problem.
They describe an algorithm that runs in $n^2 \varepsilon^{-O(\log n)}$-time, which for a given $\varepsilon>0$ produces coordinated solution paths whose joint length is at most $1+\varepsilon$ of the optimal.

The shortest joint paths (MinSum) is one of several frequently used optimization criteria in multi-robot motion planning. Additional measures, which we state here for teams of many robots (not limited to two), include
\begin{description}
    \item[MinMax:] Minimizing the length of the longest path taken by a robot in the team.
    \item[Makespan:] Minimizing the time it takes to complete the motion of all the robots in the team, namely the elapsed time from the moment the first robot leaves its start configuration to the time the last robot reaches its goal configuration.
    Here (and in the FlowTime version below) we assume that each robot moves with at most unit speed and that the robots can instantaneously change speed and direction.
    \item[FlowTime:] Minimizing the sum of the individual arrival times of the robots, namely the time it takes each robot to complete its motion, from the joint starting time.
\end{description}

We emphasize that in the setting with no obstacle, only the MinSum variant is characterized~\cite{DBLP:journals/corr/KirkpatrickL16} but not the other measures.

\begin{quote}
	\textbf{Problem~\TwoDiscsNoObstacles}: Given two unit discs moving in the empty plane (in the absence of obstacles), characterize the joint motion of the discs that minimizes the MinMax, Makespan or FlowTime.
\end{quote}

Often in robotics we wish to account for several optimization criteria simultaneously. A typical goal is to have short paths and at the same time keep large (or at least sufficient) \emph{clearance}. The clearance of a robot $R$ at a fixed configuration $q$ is the shortest distance between $R(q)$ and the obstacles. In the multi-robot case, the clearance of a robot $R_i$ in the team is the shortest distance of $R_i$ (at a fixed configuration) to the obstacles or to any other robot at the same fixed configuration (which now specifies the placements of all the robots).  

For a point robot moving among polygonal obstacles in the plane (which through employing the Minkowski sum operation applies also to a polygon translating among polygons in the plane), we can scalarize the two criteria into a single measure. Let $\pi$ denote the path traversed by the point. Parameterize the path by its length $\ell$. For every point $q\in \pi$, let $c(q)$ be the clearance at $q$, namely the shortest distance from $q$ to the obstacles. Then the cost of the path is defined as follows
$$
L(\pi)=\int_{q\in\pi} \frac{1}{c(q)} \,dq
\;.
$$

This measure has been introduced by Wein, van den Berg and Halperin~\cite{DBLP:journals/ijrr/WeinBH08}, who also gave an approximation algorithm to find the path.
The algorithm was later improved by Agarwal, Fox and Salzman~\cite{DBLP:journals/talg/AgarwalFS18}.

In preparation for the case of two discs, we write the measure $L(\pi)$ in an equivalent but slightly different manner.
Let $\pi:[0,1]\longrightarrow\mathbb R^2$ be a parameterization of the path traversed by the point, and write $\pi=(x,y)$ for the coordinate functions.
For every point $q$, let $c(q)$ be the clearance at $q$, namely the shortest distance from $q$ to the obstacles. Then the cost of the path is defined as follows
$$
L(\pi)=\int_0^1 \frac{\sqrt{{\dot x}(t)^2+{\dot y}(t)^2}}{c(\pi(t))} \,dt
\;.
$$

When we work with more than one robot then the clearance should be kept among the moving robots as well.
The following problem is an extension of Problem~\TwoDiscsMinSum to account for the combined length-clearance cost measure. 
Let $\pi=(\pi_0,\pi_1)$  be a path in ${\mathbb R}^4$. For every point along the path $(x_0,y_0,x_1,y_1)$, the first (respectively, last)  two coordinates represent a point in the path of the first (respectively, second) robot.

\begin{quote}
	\textbf{Problem~\TwoDiscsMinCombinedMeasure}: 
  Given two discs moving in the plane among polygonal obstacles with a total of $n$ vertices, devise an efficient algorithm to find (or approximate) a coordinated motion plan $(\pi_0,\pi_1)$ for the discs from free start to free target configurations such that the total combined measure is minimized, 
\begin{align*}
L(\pi_0,\pi_1) = 
 \int_0^1 \left(\sum_{i=0}^1
 \frac{\sqrt{{\dot x_i}(t)^2+{\dot y_i}(t)^2}}
      {\min(c(\pi_i(t)), \Vert \pi_i(t)-\pi_{1-i}(t)\Vert-2 )} \right) \, dt,
\end{align*}
where $c(q)$ measures the minimal distance between the robot at configuration $q$ and the obstacles, and $\pi_i:[0,1]\longrightarrow\mathbb R^2$ is a parameterization of the path of disc $i\in\{0,1\}$ from its start to target position, with coordinate functions $\pi_i=(x_i,y_i)$.
 Otherwise, if is no such a motion exists, the algorithm stops and notifies so.
\end{quote}

We subtract $2$ from $\Vert \pi_i(t)-\pi_{1-i}(t) \Vert$ because the discs collide when the distance between the centers is less than $2$.
As a warm-up, we can also consider the problem with point-shaped robots instead of discs, which eliminates subtracting $2$.
That version may have a simpler behaviour.
It is furthermore clear that there is always a solution if there exist paths from the start configurations to the targets.

The combined measure of path length and clearance together with several variants of it have been studied in the context of \emph{sampling-based robot motion planning}; see, e.g.,~\cite{choset2005principles,lavalle2006planning,DBLP:journals/cacm/Salzman19}.  A method called hybridization graphs (HGraphs for short)~\cite{DBLP:journals/trob/RavehEH11} takes as input a small number of  paths, each produced by a randomized planner for the same motion planning problem. It then adds  collision-free bridges between pairs of configurations along the paths (either a single path or two paths). Finally, it searches for the best path according to some scalarized objective function in this augmented structure. It typically produces much higher quality paths than the initial input paths, and also higher quality paths than are produced when we give a sampling-based planner (say RRT) as much time as the entire HGraphs procedure requires. This approach seems to select the best portions from each path. However it comes with no guarantees on the quality of the final solution. It would be interesting to theoretically analyze its performance, to explain its favorable behavior.

There are many interesting combinations of optimality criteria and we conclude with pointing out a combination suitable for teams of robots. Two prevalent quality measures for the motion of teams robots, which we have already mentioned for two discs, are MinSum and Makespan. It would be desirable to take these two crietria simultaneously into account in optimizing multi-robot motion planning.

\section{Can one polyhedron cage another?}

Let $A$ and $B$ be two rigid bodies (objects) in the plane or in $3$-space.
We assume that object $A$ is static and object $B$ can move. We say that object $A$ \emph{cages} object $B$ at configuration $c$ (like before we denote the object $B$ at configuration $c$ by $B(c)$) if (i) $A$ and $B(c)$ are interior-disjoint, and (ii) $B$ cannot be moved arbitrarily far from configuration $c$ without colliding with $A$.

Caging~\cite{DBLP:conf/icra/RimonB96} has been proposed and studied in robotics as a relaxed form of \emph{grasping}~\cite{Prattichizzo2016Grasping,rm-tmrg-19,DBLP:journals/ijrr/RodriguezMF12}.  Object $A$ can be viewed as a static snapshot of a robot gripper, and it holds the work-piece~$B$ in a way that, for the task at hand, sufficiently restricts the motion of object~$B$.
Consider the survey by Makita and Wan~\cite{surveyCaging2017} on robotic caging for, among others, the relation of caging to a variety of mathematical problems.

We will focus our discussion of the caging problem on polygonal or polyhedral objects. For full rigid-body motion in space, our caging problem can be formulated as follows.

\begin{quote}
	\textbf{Problem~\ThreeDCaging}
 Let $A$ and $B$ be two polyhedra in $\Reals^3$, with $m$ and $n$ vertices respectively. Efficiently determine whether $A$ can cage $B$. If so, efficiently find such a caging configuration.
\end{quote}

Thus \textbf{Problem~\ThreeDCaging} comprises two subproblems: A decision problem and the problem of actually determining a caging configuration, if one exists. As we shall argue below, the decision subproblem may in some cases be easier than finding a caging configuration. In general, though, we currently do not know to solve the decision problem more efficiently than finding a caging configuration. The statement of the problem emphasizes efficiency, since we have a polynomial time solution for it and for all the problems that we mention in this section. However, this problem raises a curious issue that is long time unresolved and leaves a notable gap between the upper and lower bounds on the running time of the algorithms for its solution. This intriguing phenomenon already occurs when the two objects are polygons in the plane, and where $B$ is only allowed to translate.

\begin{quote}
	\textbf{Problem~\TwoDCagingTrans}: Let $A$ and $B$ be two polygons in $\Reals^2$, with $m$ and $n$ vertices respectively. Polygon $B$ can only translate. Efficiently determine whether $A$ can cage $B$ under translation only. If so, efficiently find such a caging configuration.
\end{quote}

We now take a small side-track and define a fundamental mathematical operation, which is ubiquitous in robotics and automation as well as in CAD/CAM, 3D modeling, layout optimization, and even economics: \emph{Minkowski summation}. For two sets $P,Q$ in $\Reals^d$, their Minkowski sum, denoted $P\oplus Q$, is the pairwise vector sum of the points in both sets, that is $P\oplus Q=\{p+q|p\in P,q\in Q\}$. The Minkowski sum of two polygons is a polygonal set.
Let $-P$ denote the object $P$ in $\Reals^d$ reflected through the origin, namely $-P=\{-p|p\in P\}$. 
We demonstrate the connection between the caging problem and Minkwoski sums by the following equivalent statement of \textbf{Problem~\TwoDCagingTrans} (see \Cref{MinkowskiSumPolygons}):

\begin{figure}
\centering
\includegraphics[page=1]{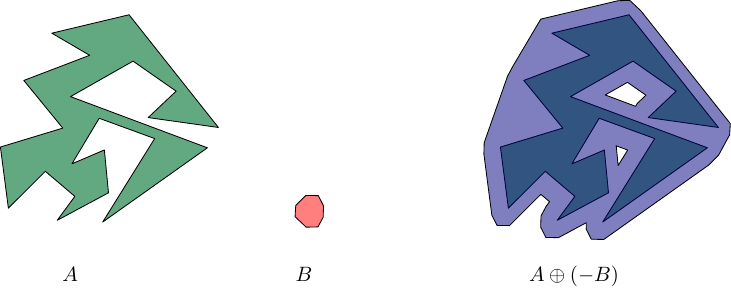}
\caption{Polygons $A$ and $B$ and the Minkowsky sum $A\oplus (-B)$ on top of $A$.
As is seen, $A\oplus (-B)$ has two holes, which correspond to the two ``pockets'' of $A$ that cage $B$.}
\label{MinkowskiSumPolygons}
\end{figure}

\begin{quote}
	\textbf{Problem~\TwoDHoleInMinkowskiSum (equivalent to~\TwoDCagingTrans)}: Let $A$ and $B$ be two polygons in $\Reals^2$, with $m$ and $n$ vertices respectively. Does the Minkowski sum $A\oplus (-B)$ contain a hole? If so, find a point inside this hole.
\end{quote}

The equivalence between the two problems stems from the following simple observation.
\begin{observation}
Let $A$ and $B$ be two rigid bodies in $\Reals^d$, and $t$ a vector in $\Reals^d$. The object $B$ translated by the vector $t$, namely $B+t$, intersects $A$ if and only if $t\in A\oplus -B$.
\end{observation}

The vector $t$ represents the configuration of the moving object (robot) $B$ when only translations are allowed. Therefore the configuration space of this motion problem is $\Reals^d$ and 
$A\oplus -B \subseteq \Reals^d$ is the collection of forbidden configurations, namely, configurations where the translated $B$ intersects with $A$. In caging, we are looking for a translation $t$ such that $t\notin A\oplus -B$. However, this translation $t\in \Reals^d$ needs to be completely surrounded by points in $A\oplus -B$, as otherwise there will be a path from $t$ to infinity in the complement of $A\oplus -B$ and $A$ will not cage $B$. In other words $t$ must be inside a hole in $A\oplus -B$.

Let's assume for simplicity that $m=n$. We can compute the Minkowski sum $A\oplus -B$ in  $O(n^4)$ time. This algorithm in turn will reveal whether there is a hole in the sum and will also give us a caging configuration when there is a hole in the sum. However, in general we do not know to decide whether there is a hole in the Minkowski sum of two polygons with $n$ vertices each in $o(n^4)$ time. Erickson~\cite{Erickson2016MinkLowerBound} showed that the 
\textbf{Problem~\TwoDHoleInMinkowskiSum} is so-called \emph{$6$-SUM hard}.
In the problem $k$-SUM, we are given integers $x_1,\ldots,x_n$, and we want to know if there exist $k$ distinct of these integers that sum to $0$.
It is folklore that this problem can be solved in $O(n^{\lceil k/2\rceil})$ time~\cite{DBLP:conf/icalp/AbboudL13}.
In certain restricted models of computation, a lower bound of $\Omega(n^{\lceil k/2\rceil})$ has also been shown~\cite{DBLP:journals/jacm/AilonC05,cj99-08}.
It has therefore become a popular conjecture that $k$-SUM cannot be solved in $O(n^{\lceil k/2\rceil-\varepsilon})$ time for any $\varepsilon>0$ in the standard word RAM model.
Assuming this conjecture, the $6$-SUM hardness implies that there does not exist an algorithm for \TwoDHoleInMinkowskiSum with running time $O(n^{3-\varepsilon})$.
This still leaves a gap in the order of magnitude of $n$ between the lower bound and the running time of the existing algorithm. 

For the first problem of this section, \textbf{Problem~\ThreeDCaging}, we need to generalize the notion of Minkowski sums, and resort to the general theory of arrangements of surfaces~\cite{hs-a-18}. By standard arguments in this theory, \textbf{Problem~\ThreeDCaging}, assuming $m=n$, can be solved in $O(n^{16+\varepsilon})$ time~\cite{DBLP:journals/tcs/ChazelleEGS91,DBLP:journals/jacm/Koltun04}, for any fixed $\varepsilon>0$. Improving the running time in this case introduces many challenges, and some of them are fundamental and hard questions in the study of arrangements of surfaces in higher dimensions and are beyond the scope of this survey. So let's get back to the two-dimensional translational case.

Finding an efficient algorithm for \textbf{Problem~\TwoDCagingTrans}, or equivalently for \textbf{Problem~\TwoDHoleInMinkowskiSum}, seems a hard nut to crack. What else can be done? Karasev~\cite{DBLP:journals/dcg/Karasev10b} has characterized a family of non-convex polygons with the property that the Minkowski sum of any pair of them does not have holes. Note that this is trivial for convex polygons, as the Minkowski sum of a pair of convex polygons is a convex polygon. Karasev suggests a measure for how far the summand polygons are from being convex, and if they are not too far by that measure, then their Minkowski sum does not have holes, which can be decided much more efficiently than constructing the sum. Can we find other families of polygons for which the Minkowski sum of pairs of them is guaranteed not to have holes?

Being a ubiquitous operation in many applications, and since the Minkowski sum of two polygons with a total of $n$ vertices can have very high (quartic) complexity, it is essential to be efficient in computing the sum. In particular, we are in search of an output-sensitive algorithm.  Erickson's result~\cite{Erickson2016MinkLowerBound}  implies that we will probably not be able to find such an algorithm with less than an $\Omega(n^3)$ running-time overhead. Thus we seek alternative ways to speed up this computation.
Baram, Fogel, Halperin, Hemmer and Morr~\cite{DBLP:journals/comgeo/BaramFHHM18} studied a problem on holes in the summands of the Minkwoski sum: They showed that if the summands of the Minkowski sum operation are polygons possibly with holes, then there are certain easy-to-apply rules, which will always eliminate the holes from at least one of the summands, without affecting the resulting Minkowski sum. The elimination of holes makes the construction of the Minkowski sum when some holes are ``filled'' more efficient in practice; see experiments in~\cite[Section~4]{DBLP:journals/comgeo/BaramFHHM18}. Can we find other rules that are efficient to apply and will make the computation of the Minkowski sum faster?

Minkowski sums raise a wealth of absorbing (open) problems in combinatorics and computation. We conclude this section with the issue of inverse problems in Minkowski sums or \emph{Minkowski sum deconstruction}: Given a set $M \in \reals^d$, is $M$ the Minkowski sum (or its approximation) of two non-trivial\footnote{By non-trivial we preclude the case where one summand is a point.} summands in $\reals^d$?
For motivation and examples, see~\cite{DBLP:journals/dcg/BerberichHKP12}.

\section{Interpolate between easy and hard instances of multi-robot motion planning}

Multi-robot motion planning (MRMP, for short) is one of the most intriguing and challenging problems, both from the robotics perspective as well as from the computational geometry perspective. The recent years have seen an increasing global trend to use fleets of robots for logistics, inspection and surveillance tasks, agriculture, maintenance and repair, and more. New and demanding applications often require sophisticated, ultra fast coordination. At the same time, multi-robot motion planning has been proved to be \emph{computationally hard}\footnote{For a general review of the computational complexity of motion planning, see, e.g., the survey by Solovey~\cite{Solovey2020}.}~\cite{hss-cmpmio,SolHal16j,sy-snp84}. 

MRMP is also highly varied, and often means different things to different people. For example, there is a vast body of work on the discrete case of graphs---Multi-Agent Path Finding (MAPF)~\cite{MAPFALGOREVIEWGAO2024, MAPFReview2023}, or swarm solutions, which often follow a distributed approach~\cite{SWARMREVIEW2023}. Here we deal with a basic continuous variant, where there is a central control that knows the environment and the exact location of each robot at all times.

\begin{figure}
\centering
\includegraphics[page=9]{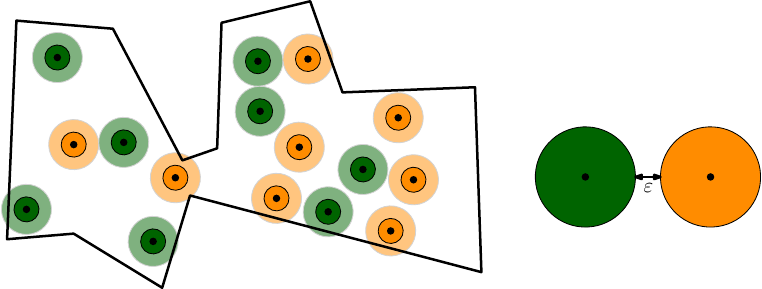}
\caption{Left: The setting solved in ~\cite{DBLP:journals/tase/AdlerPSG15}:
The dark green and orange discs are the start and target positions, respectively.
The light colored larger discs of radius $2$ show that the distance between any pair of start and/or target positions is at least $4$.
Right: We propose to study the situation where the distance is at least $2+\varepsilon$ for an arbitrary value $\varepsilon>0$.}
\label{fig:discInterpolate}
\end{figure}

Even this fundamental variant is far from being completely understood.
While the problem is in general hard, various scenarios are relatively easily solvable.
We focus on a specific MRMP problem, where we know to distinguish between easy and hard instances, and for which we have an efficient solution tailored for  easier instances. The setting is as follows. There are $m$ unit disc robots moving in a planar workspace, which is a single simple polygon with $n$ edges. The robots are located at their given start positions, and we are also given $m$ target positions. The problem is \emph{unlabeled} in the sense that we do not care which robot goes to which target position, as long as at the end of the motion each target position is occupied by one robot. We specify each start or target position by the coordinates of the center of the robot at that position. We  impose one extra condition: We require that between every pair of start or target position there is a distance of at least 4 units. For this setting there is always a solution and there is an algorithm with running time $O(n\log n+mn+m^2)$ to find such a solution~\cite{DBLP:journals/tase/AdlerBHS15}; see \Cref{fig:discInterpolate} (left) for an example of an instance. This result has later been sharpened in allowing tighter separation conditions between start and target positions, while still always having a solution~\cite{DBLP:conf/compgeom/BanyassadyBBBFH22}.

If we do not assume the extra spacing condition then the problem is presumably PSPACE-hard, even though we are in the unlabeled setting.
This was proved for unit squares of polygons with holes~\cite{SolHal16j}, and for disc robots of two different sizes~\cite{DBLP:conf/fun/BrockenHKLS21}. 
More recently it has also been shown for unit squares inside a simple polygon and unit discs in a polygon with holes~\cite{abrahamsen2024reconfiguration}.

The question we pose is to devise an algorithm that smoothly interpolates between `easy' instances as above and hard instances where such separation is not assumed. We introduce another parameter to the problem, $\varepsilon$, which measures how much above 2 is the separation between any pair of start and/or target positions; see \Cref{fig:discInterpolate} (right).
We assume that $\varepsilon>0$, namely that robots placed at the pair of positions that achieve the minimum distance are not osculating.

\begin{quote}
    \textbf{Problem~\UnlabeledMultiDiscInterpolate}
    A collection of  $m$ unlabeled unit discs are moving inside a simple polygon with $n$ edges. We are given \emph{valid} start and target  positions for the $m$ discs, namely the robots at their start (respectively target) positions are pairwise interior-disjoint. Let $d:=2+\varepsilon$ denote the minimal distance between the start and target positions of the robots for a value $\varepsilon>0$.
    Devise an algorithm to find a collision free motion plan for the robots from start positions to target positions with running time that is polynomial in $m,n$, and $1/\varepsilon$, or decide that it does not exist.
\end{quote}

In this context we mention the results of Alt, Fleischer, Kaufmann, Mehlhorn, N{\"{a}}her, Schirra and Uhrig~\cite{DBLP:journals/algorithmica/AltFKMNSU92} on motion planning for a rectangular robot translating and rotating among polygonal obstacles in the plane. This work stands out in the motion planning literature in computational geometry in that it introduced the \emph{tightness} of the workspace as a measure of the difficulty of the problem. It also analyses the running time of the algorithm both as a parameter of the complexity of the obstacles as well as of this tightness, which is similar in this respect to the problem we have just posed (and different from it in various other ways).

Finally, we mention several recent works by Geft et al., which aim to delineate the boundary between easy and hard problems in MAPF and MRMP, typically by considering special properties (for example, monotonicity---see Section~\ref{sec:reconfig}) or certain extra parameters, beyond separation, that arise in these problems~\cite{DBLP:conf/atal/AbrahamsenGHU23,DBLP:conf/socs/Geft23,DBLP:conf/atal/GeftH22,GeftWAFR2024}.

\section{Geometric reconfiguration and object rearrangement}\label{sec:reconfig}

In a \emph{valid configuration} of a set of objects in two- or three-dimensional space, each object has a fixed pose (i.e., position and orientation) and the objects are pairwise interior-disjoint. In \emph{object reconfiguration} we are given two valid configurations of the same set of objects, a start configuration and a target configuration, and the goal is to plan collision-free moves that will take the objects from start to target, or determine that no such reconfiguration is possible.
Naturally, numerous tasks in robotics and automation can be cast as object reconfiguration, including packaging, product assembly,  product maintenance, shelf stocking,  to mention just a few.

The family of reconfiguration problems is broad and has an abundance of variants and occurrences.
We distinguish between two main categories in this family: (i) \emph{basic} object reconfiguration, where the objects are ``free-flying'', and (ii) \emph{manipulated} object reconfiguration, where the objects are moved from start to target by robots. The basic category is relevant when the objects have autonomy of motion, when they can be manipulated without direct contact (e.g., using force fields~\cite{BoehringerPhDThesis1997}), or as a preprocessing step to certain solutions of the typically much harder manipulated object reconfiguration. The problems of this basic type have been researched for decades in mathematics, computer science, and various engineering disciplines~\cite{DBLP:journals/corr/cs-DM-0204002,DBLP:journals/comgeo/DumitrescuJ13}.    
Interest in problems of the manipulated type is on the rise with the increasing usage of robots in ever more complex task; see, e.g., \cite{HuangZhangYuICRA2024,10161528,DBLP:journals/tase/ShomeSYBH21,DBLP:conf/corl/TangS22}.

How is object reconfiguration different from multi-robot motion planning? These are two largely different families of problems, but they are similar when we discuss free-flying object reconfiguration with complex allowable motions.
The term \emph{reconfiguration} has been used in the context of planning the motion of many moving bodies, when the type of motion allowed is very simple~\cite{DBLP:journals/comgeo/DumitrescuJ13}; these are sometimes referred to as coin-moving puzzles~\cite{DBLP:journals/corr/cs-DM-0204002}. Reconfiguration algorithms often aim to optimize the overall effort exerted in moving the objects from one configuration to another, typically in the absence of obstacles and using simple allowable motions. It is reminiscent of shape formation problems, which have been studied in robotics~\cite{DBLP:journals/ijrr/Alonso-MoraBR17,Jeon2014MultiRobotFS}.
Other objectives have likewise been considered, such as \emph{space-aware} reconfiguration, where the goal is to minimize the physical space needed for the reconfiguration to be possible~\cite{DBLP:journals/dcg/HalperinKMS23}, or makespan as defined in Section~\ref{sec:shortest-coordinated} (see, e.g., \cite{DBLP:journals/siamcomp/DemaineFKMS19}).

\begin{figure}
\centering
\includegraphics[page=12]{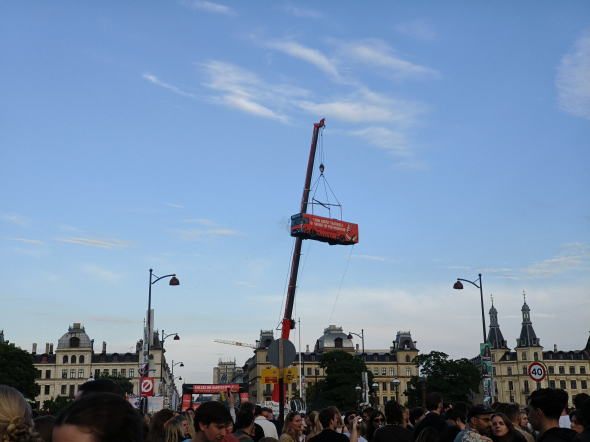}
\caption{The lifting model demonstrated during the festival \emph{Copenhagen Distortion}, 2024.
Picture by Hao Wu.}
\label{fig:lifting}
\end{figure}

Many restrictions on the motions can be considered.
Three natural models, for objects in the plane, are (i) the \emph{lifting} model, where each move consists of lifting an object (in the third dimension, say) and placing it at a free position (see \Cref{fig:lifting}), (ii) the \emph{sliding} model, where the objects can slide along arbitrary continuous curves, and (iii) the \emph{translating} model, in which the objects can move along straight line segments.
Dumitrescu and Jiang~\cite{DBLP:journals/comgeo/DumitrescuJ13} showed that for unlabeled unit discs, it is NP-hard to find the minimum number of moves (with one object moving at a time) to reconfigure from the start positions to the target positions in the sliding and translating models.
Another restriction often studied in reconfiguration is \emph{monotonicity}, where each object is allowed to move exactly once, from its start to a target position.

There is a simple greedy algorithm for finding a suitable sequence of moves if it exists in the monotone lifting model:
We say that a target is \emph{finished} if it coincides with a disc.
A target is \emph{blocked} if it is not finished and there is a disc overlapping it partially.
A target is \emph{available} if it is neither finished, nor blocked.
Note that we can only move a disc to an available target.
If all targets are finished, we are done.
Otherwise, we give up if there is no available target.
If there is an available target, we find the blocked target that is blocked by a minimum number of discs and move one of them to an available target.
This makes sure that we minimize the number of moves before we get one more available target.

The situation becomes more complicated and interesting when we consider the sliding or translating model, which leads us to the following questions.

\begin{quote}
\textbf{Problem~\UnlabeledMonotoneReconfiguration}
Consider the following question:
Given two sets $S$ and $T$, each consisting of $n$ pairwise interior-disjoint unit discs in the plane (but discs in $S$ can overlap discs in $T$), is there a monotone reconfiguration from $S$ to $T$ using sliding moves?
Can the problem be solved in polynomial time?
\end{quote}

As a warm-up, one might consider the case of $2\times 2$ squares placed on the integer grid, possibly even restricting the motions to horizontal and vertical translating moves.
Even in this case, the problem does not seem to be trivial.

In \emph{manipulated object reconfiguration}, where we also take into account the robots that move the objects, further complications arise. Already for one robot, we need to make sure that the robot does not collide with the objects that are not expected to move at a certain stage of the process. When two or more robots simultaneously reconfigure the objects, we are faced also with multi-robot motion coordination (see, e.g., \cite{DBLP:journals/tase/ShomeSYBH21}). Then, we may additionally have the flexibility to choose preferred placement of the bases of the manipulating robots, an optimization that could significantly simplify or speed up the task. Due to the complexity of each of these sub-problems they are often 
tackled in separation, which may result in far from optimal solutions. While taking them all into account simultaneously is in general prohibitively difficult, it would be interesting to discover settings where this is feasible and can lead to a complete optimized solution for the ``full manipulated-reconfiguration stack.''

\section{Interlocked objects with fewer convex parts}\label{sec:interlocked}

Assembly planning is a large sub-area of robotics where the aim is to move a set of objects from initial positions to a desired target configuration without causing the objects to collide.
It is often conceptually easier to study the equivalent reverse question, namely whether objects that start in a specific configuration, typically as a dense cluster, can be moved apart using prescribed types of motions.

Since assembly in practice is often performed by adding one object after the other, a variant of particular importance is whether we can move one object away from the rest at a time.
If the objects are not convex, it is easy to come up with examples of two objects that are \emph{locked} into each other, so that they cannot be moved apart.
This is even the case in the plane, where the objects can for instance be modelled as polygons.
However, if we are dealing with \emph{convex} objects in the plane, there always exists an object $P$ that is not blocked from, say, above (namely in the $y$ direction) by another object~\cite{DBLP:conf/stoc/GuibasY80}, so $P$ can be moved vertically up without collisions, and this process can be repeated until all objects have been moved away from each other.

\begin{figure}[h] \centering
\includegraphics[page=13]{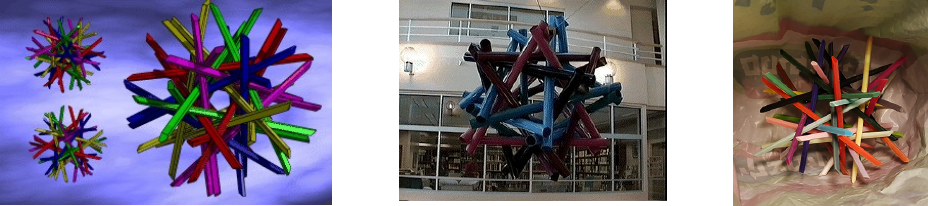}
    \label{fig:example_instance}
    \caption{The Snoeyink-Stolfi construction of 30 convex parts that cannot be taken apart with two hands. Left: An illustration of the actual construction. Middle: A sculpture at UBC inspired by the construction. The left and middle figures are taken from Snoeyink's website \url{https://www.cs.unc.edu/~snoeyink/}.
   Right: The construction printed on a  plaster-based 3D printer at Tel Aviv University.}
    \label{fig:Snoeyink30}
\end{figure}

This is not the case in the three-dimensional space:
Three cylinders can block each other cyclically so that none of them can be translated vertically up.
Snoeyink and Stolfi~\cite{DBLP:journals/dcg/SnoeyinkS94} showed that in any configuration of at most five convex objects in space, there is a direction $\vec{d}$, so that an object can be translated away from the others in direction $\vec{d}$.
They also gave an example of six objects so that there is no such direction.
Even stronger, it is not possible to partition the objects into two groups, where one group can be moved away from the other by translation.
In other words, the objects cannot be taken apart by two \emph{hands}, if these hands can only translate and not rotate the objects.
However, if arbitrary rigid motions are allowed, that is, the objects can also be rotated during the movement, then two hands suffice.
Snoeyink and Stolfi then described a configuration of $30$ convex objects that cannot be moved away from each other by two hands with arbitrary rigid motions allowed. See Figure~\ref{fig:Snoeyink30}.
The configuration is completely locked: There exists no subset that can be moved even infinitesimally without causing a collision with the remaining objects.

Additional interesting inseparability results of this type have been shown over the years; see, e.g., the review by Toussaint\footnote{Godfried Toussaint, who passed away in 2019, presented a more modern review of similar problems in the International Conference on Mathematics and Mechanics in Nice in 2017. To obtain the accompanying mini-survey that he wrote for the occasion, contact the second author at \texttt{danha@tauex.tau.ac.il}.}~\cite{Toussaint85}. 

Using a hand for each object, it is possible to disassemble any configuration of convex objects $P_1,\ldots,P_n$:
Choose a reference point $p_i$ in each object $P_i$, and translate $P_i$ by the velocity vector $p_i-p_1$.
This corresponds to scaling up the initial configuration and then scaling down each object $P_i$ to its original size, keeping the reference point $p_i$ fixed.
Hence, the movement does not cause collisions (this in fact works for any collection of star-shaped objects, where the reference points should be chosen from the respective kernels of the objects).
Let us conclude this section by suggesting some interesting questions for future research.

\begin{quote}
\textbf{Problem~\ThreeInterlockeddFewerThanThirty}:
Does there exist a configuration with fewer than $30$ convex objects that cannot be taken apart by two hands?
\end{quote}

How many hands may be needed to move apart $n$ convex objects, as a function of $n$?
All we know is that $2$ do not always suffice and $n$ are always sufficient.
For a given configuration of objects, is there an algorithm for deciding how many hands are needed and describing how to move the objects apart?

To show that their construction of 30 objects cannot be taken apart with two hands, Snoeyink and Stolfi considered an exponential number of decompositions of the objects into two proper subsets, and for each such decomposition applied linear programming to test whether one subset can be moved  infinitesimally from the other~\cite{DBLP:journals/dcg/SnoeyinkS94}. Later, Guibas, Halperin, Hirukawa, Latombe and Wilson~\cite{DBLP:journals/ijcga/GuibasHLW98} gave a polynomial time algorithm for the same problem, which runs in time that is roughly $O(n^{12})$, where $n$ is the complexity of the polyhedral parts.\footnote{The actual running time is governed by parameters related to the kinematic analysis of the contacts between the parts under infinitesimal rigid motion; see~\cite{DBLP:journals/ijcga/GuibasHLW98} for details.} Can one devise a more efficient algorithm for the task? This seems highly plausible by employing sophisticated data structures. 

We remark that in assembly planning one often has to take into consideration additional requirements, beyond  avoiding collision between parts. For example, a practical requirement is that the two sub-assemblies, held by the two hands, will each constitute a connected object. Interestingly, the connectivity requirement makes the problem computationally hard even in very simple settings~\cite{DBLP:conf/soda/AgarwalAGH21}.

\section{Minimal design modification, or explainability}\label{sec:MDM}

In everyday life as well as in many industrial settings, it is often the case that it is impossible to perform some desired task because of unfavorable properties of the scenario.
We then seek to change the setting just enough to be able to carry out the task. Another view of the same problem is, our algorithm fails to find a feasible plan for the task, and we would like to give the user an explanation (or an excuse, as it is called in~\cite{DBLP:conf/aips/GobelbeckerKEBN10}), why the algorithm failed; in such a case a compact or small explanation is desirable~\cite{DBLP:journals/ijrr/Hauser14}. 
We use the term \emph{minimum design modification} as an umbrella term for problems belonging to this category, where the modification is geometric such as ``slightly modify the pose of an object in the input'', ``remove an obstacle'', or ``change the dimensions of an object''.
A minimum design modification problem should consist of (i) a desired task/plan, (ii) a set of allowable modifications, and (iii) an objective function.

\begin{figure}
\centering
\includegraphics[page=3]{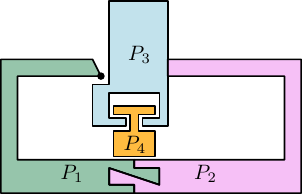}
\caption{This configuration cannot be disassembled with two hands, but moving the marked vertex of $P_1$ a bit to the left, the polygons $P_3$ and $P_4$ can be moved up and away from $P_1$ and $P_2$.}
\label{fig:MDM1}
\end{figure}

As introduced in \Cref{sec:interlocked}, assembly planning is concerned with designing a sequence of motions that will bring a set of separate parts into their desired juxtaposition in a final product.
In general the problem is hard and we add constraints to make it tractable.
One such constraint is the number of hands used to move the pieces (see \Cref{sec:interlocked,sec:hands}), and another one is the allowable motions of the parts.
We assume here that the assembly must be carried out by two hands and by translating the parts.
Let us again explain what this means in the context of \emph{disassembling} rather than assembling the parts:
We wish to partition the parts forming the final product into two groups so that one group can be moved, as a single rigid object, arbitrarily far away by translating them together along a single line segment, without colliding with the other group.
This process can be repeated recursively on the two groups so that in the end, the assembly has been split into the individual parts.
For polyhedral parts, it can be decided efficiently whether an assembly can be disassembled like this~\cite{DBLP:journals/tase/FogelH13,DBLP:journals/algorithmica/HalperinLW00,DBLP:journals/ai/WilsonL94}.
For simplicity of exposition, we restrict ourselves to two-dimensional assemblies of polygons.
Now assume that we are given an assembly that cannot be disassembled with these constraints, as for instance is shown in \Cref{fig:MDM1}.
We are also given a tolerance parameter $\varepsilon>0$, meaning that we are allowed to move each vertex by a distance of at most $\varepsilon$, as long as we avoid self-intersections.
For a more elaborate model of toleranced modification, see, e.g.,~\cite{DBLP:journals/cad/LatombeWG97}.

As our first problem, we just want the assembly to be partitioned into two parts that can be disassembled from each other:

\begin{quote}
\textbf{Problem~\MDMTwoDPartitionFiniteTrans}:
We are given a two-dimensional assembly $A$ made of polygons and a tolerance parameter $\varepsilon>0$.
What is the minimum number of vertices in the polygons that need to be moved by a distance of at most $\varepsilon$ such that the modified assembly $A'$ can be partitioned into two parts that can be moved arbitrarily far away from each other by two hands using a single translation?
Devise an efficient algorithm to answer this question.
\end{quote}

We also ask the following more general question, where we want to disassemble the assembly entirely.

\begin{quote}
\textbf{Problem~\MDMTwoDAssemblyFiniteTrans}:
As in the previous problem, but we want to be able to repeat the process so that all parts are moved away from each other.
\end{quote}

An interesting variant of \textbf{Problems~\MDMTwoDPartitionFiniteTrans} and~\textbf{\MDMTwoDAssemblyFiniteTrans}
 is to find the smallest value of $\varepsilon$ such that the assembly can be partitioned or assembled as above, namely by moving each vertex distance at most $\varepsilon$.

We now turn to a different manufacturing process. Casting is a process where liquid material is poured into a mold having the shape of a desired product.
After the material solidifies, the product is removed from the mold.
We want to remove our produced object from the mold without collision.
The casting problem is closely related to assembly planning; indeed, we need to make sure that the mold and the cast object can be disassembled from each other.

\begin{figure}
\centering
\includegraphics[page=4]{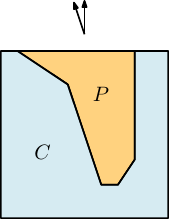}
\caption{The object $P$ can be moved away from the cast $C$ by translation.
The arrows show the range of valid removal directions.}
\label{fig:MDM2}
\end{figure}

To model the casting problem more formally, we assume that the object $P$ to be cast is a polyhedron in $\mathbb R^3$ with a boundary consisting of $n$ polygonal facets.
The mold is box-shaped and has a cavity whose shape is exactly that of $P$, such that one of the facets of $P$ is the top facet of the cavity.
See \Cref{fig:MDM2} for an illustration in 2D.
We wish to move $P$ away from the mold by translation in a single direction.
A direction that does not cause collision is called a \emph{valid removal direction}.

Bose, Geft, Halperin and Shamai~\cite{DBLP:journals/corr/abs-1708-04203,DBLP:conf/case/BoseHS17} showed that one can decide for each facet of $P$ whether $P$ is castable with that facet used as the top facet, for a total time of $O(n)$ for all the facets.
They then showed how to compute the set of all valid removal directions for all top facets in $O(n\log n)$ time and that this is asymptotically optimal.
Interestingly, they also showed that there can be at most six top facets for which the set of valid removal directions is non-empty.
These results improved earlier work on the casting problem~\cite{DBLP:journals/cad/AhnBBCHMS02,DBLP:journals/algorithmica/AsbergBBGOTWZ97}.

When no facet can be used as a top facet, we may want to modify $P$ as little as possible to another object $P'$ that \emph{is} castable, i.e., we can choose a top facet that has a valid removal direction.
This leads to our next problem, which is a special case of \MDMTwoDPartitionFiniteTrans, but in three dimensions rather than two.

\begin{quote}
\textbf{Problem~\MDMCastingWithOnePartMold}:
We are given a three-dimensional polyhedron $P$ and a tolerance parameter $\varepsilon>0$.
What is the minimum number of vertices that need to be moved by a distance of at most $\varepsilon$ such that the modified polyhedron $P'$ is castable?
Devise an efficient algorithm to answer this question.
\end{quote}

Minimal design modification naturally arises in \emph{reconfiguration problems} (see \Cref{sec:reconfig}).
Let us again consider a setting where we wish to get from one configuration of unit disc robots to another (we refer to the start and target placements collectively as \emph{endpoint} placements). 
It might be impossible because any way of moving the discs to their final positions would cause the discs to collide.
In the remainder of this section we focus on \emph{monotone} reconfiguration, where discs are moved one after the other, and in one move a disc reaches its destination where it stays thereafter. Also we assume that the scenes are obstacle free. One can clearly consider generalizations of these premises.
To make an instance solvable a possible modification is to move some endpoint placements by $\varepsilon>0$ from their original setting.

\begin{quote}
\textbf{Problem~\MDMTwoDReconfigurationUnitDiscsMove}:
Given an infeasible instance of unit disc reconfiguration (labeled or unlabeled), what is the minimum value $\varepsilon>0$ so that if all endpoint placements are allowed to move by $\varepsilon$, we arrive at a feasible instance?
Devise an efficient algorithm to answer this question.
\end{quote}

We can also ask for the minimum \emph{number} of endpoint placements that should be moved by a given value $\varepsilon$ to reach a feasible configuration. 

Alternatively, we can ask for the minimum number of robots whose radii should be shrunk by a value at most some fixed $\varepsilon>0$ so that we arrive at a feasible instance. We call this 
\textbf{Problem~\MDMTwoDReconfigurationUnitDiscsMinNumShrink}, which can be either labeled or unlabeled.
An answer to this question can be interpreted as an explanation for why the original instance is not feasible.

The minimum radius variant of this question could be relevant when the discs' radii had been uniformly enlarged to account for uncertainty and safety issues. This enlargement resulted in an infeasible instance, and we wish to find a smaller enlargement (but the largest possible) that would make the instance feasible. This could be phrased as a maximal enlargement problem, but to keep with the spirit of minimal design modification we phrase it as ``what is the minimum value $\varepsilon>0$ so that if all discs shrink by $\varepsilon$ (keeping their center points), we arrive at a feasible instance?'', and we call the latter 
\textbf{Problem~\MDMTwoDReconfigurationUnitDiscsShrink}.

\paragraph{Remark.} Note that if moving discs by $\varepsilon$ leads to a feasible configuration, then shrinking by $\varepsilon$ also does, whereas the opposite implication is not true in general.

\section{Assembly planning with more than two hands}\label{sec:hands}

We have already discussed the problem of \emph{assembly partitioning} in previous sections. For completeness, let's review it one more time: We are given a collection $A$ of pairwise interior-disjoint objects in two- or three-dimensional space, together with an allowable set of motions, and we ask whether there is a proper subset $S\subset A$ that can be moved away as a rigid body from the rest of the objects, $A\setminus S$, without colliding with the objects in $A\setminus S$. If so, we wish to output the subset $S$ and a specific collision-avoiding motion (from the allowable set).
In Section~\ref{sec:interlocked}\slash Problem~\ThreeInterlockeddFewerThanThirty,
the allowable motion are infinitesimal translations and rotations in 3-space, and in Section~\ref{sec:MDM}/Problem~\MDMTwoDAssemblyFiniteTrans, the allowable motions are infinite translations in the plane.

\begin{figure}
\centering
\includegraphics[page=6]{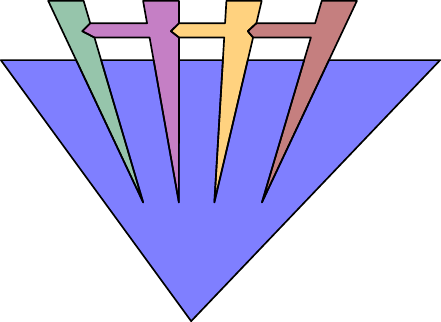}
\caption{A $5$-handed assembly, following an example from~\cite{DBLP:conf/compgeom/Natarajan89}.}
\label{fig:5hands}
\end{figure}

Both these partitioning problems are instances of $2$-handed partitioning, where we split the assembly into \emph{two} subsets, and one of them needs to move away from the other. For certain partitioning problem two hands may not suffice: Consider for example Figure~\ref{fig:5hands}. If we only allow infinite translations in the plane then we need to partition the assembly into five subsets, one stationary and each of the other moving along its distinct individual direction, and we refer to this assembly as $5$-handed.

The most prevalent industrial assembly operations are $2$-handed. Furthermore, there is a general framework~\cite{DBLP:journals/algorithmica/HalperinLW00,DBLP:journals/ai/WilsonL94} to solve assembly partitioning for constant-description allowable motion. The idea is to study the problem in the so-called \emph{motion space}, whose dimension equals the number of degrees of freedom of the path, where every point represents a possible path in the family of allowable motions. This space is then overlaid with an arrangement~\cite{hs-a-18} of certain critical hyper-surfaces, such that inside each cell of this arrangement all the paths corresponding to points in the cell induce the same valid partitions. 

We say that an assembly $A$ is \emph{$k$-handed}, for $k>2$, if it cannot be partitioned into $k-1$ or fewer subsets that are separable as above, but it can be so partitioned with $k$ hands. 

We look for an extension of the motion-space framework to $k>2$-handed assembly partitioning. It would already be interesting to solve this for $3$ hands for infinite translations in the plane.

\begin{quote}
\textbf{Problem~\ThreeHandedPartitionInfiniteTrans}:
Given an assembly $A$ of polygons in the plane, which we assume not to be 2-handed, devise an efficient algorithm to find two 
proper subsets $A_1,A_2\subset A$, such that $A\setminus (A_1\cup A_2)$ is non-empty, together
with two vectors $d_1,d_2$, such that $A_1$ and $A_2$ can be moved simultaneously: $A_1$ along the direction $d_1$ at speed $|d_1|$ and $A_2$ along direction $d_2$ at speed $|d_2|$, such that throughout the motion the three sets $A_1,A_2$ and $A\setminus (A_1\cup A_2)$ are pairwise interior-disjoint. Otherwise, report that no such motion exists.
Devise an efficient algorithm to answer this question.
\end{quote}

Naturally, we are  interested in extensions of
Problem~\ThreeHandedPartitionInfiniteTrans not only to $k>3$ as mentioned, but also to more complex settings of objects and allowable motions.

\section{Compact representation of spatial objects}

Polyhedra are three-dimensional objects, and they are used in robotics to model the actual robots and obstacles in the environment.
They are likewise used in several other practical domains such as computer graphics, engineering and simulation to model all kinds of spatial objects.
Mathematically, we can define a polyhedron as a compact region in $\mathbb R^3$ bounded by a finite number of (two-dimensional) polygons.
There are two main types of representations used to model a polyhedron $P$.
The perhaps simplest and most frequently used is to just specify the surface of $P$ as a collection of triangles.
This is known as a \emph{surface} model.
Knowing the surface, we also implicitly know the volume inside $P$.

In some cases, however, it is convenient to have an explicit representation of the interior of $P$ and not only the surface.
To this end, we can use a \emph{tetrahedralization} $\mathcal T$ of $P$.
Here, $\mathcal T$ is a collection of tetrahedra 
 whose union is $P$, and any two tetrahedra in $\mathcal T$ are either disjoint or intersect in a face of both of dimension at most $2$ (i.e., the intersection is a vertex, an edge or a (triangular) facet of both tetrahedra).
This is a so-called \emph{volumetric} model, since the tetrahedra together consistute the entire volume of $P$.

Tetrahedralizations are also called \emph{triangulations}, and they are indeed three-dimensional analogous to triangulations of two-dimensional polygons.
A surface model of a polyhedron corresponds to representing a polygon just by its boundary edges, whereas a volumetric model (using a tetrahedralization) corresponds to representing a polygon by a triangulation.
However, whereas triangulations of polygons are very well understood, many mysteries surround their three-dimensional counterparts.

A fundamental task is to convert a model of one type into the other.
Obtaining a surface model from a volumetric model is quite straight-forward:
We just need to identify the facets of tetrahedra where there is no tetrahedron on the other side.
Such facets must necessarily belong to the surface of $P$, so they together constitute a surface model of $P$.
However, obtaining a volumetric model from a surface model is far more challenging and some prominent research questions remain to be answered in this direction.

An important property of a representation is whether or not it uses so-called Steiner points.
A \emph{Steiner point} is a vertex of a triangle/tetrahedron that is not a vertex of the object $P$.
Any simple polygon with $n$ vertices can be triangulated without Steiner points, and all such triangulations consist of $n-2$ triangles.
Any \emph{convex} polyhedron with $n$ vertices can be tetrahedralized without Steiner points, but different tetrahedralizations may have different numbers of tetrahedra, and the number can vary from $\Theta(n)$ to $\Theta(n^2)$~\cite{BernEppstein1995}.

\begin{figure}
\centering
\includegraphics[page=8]{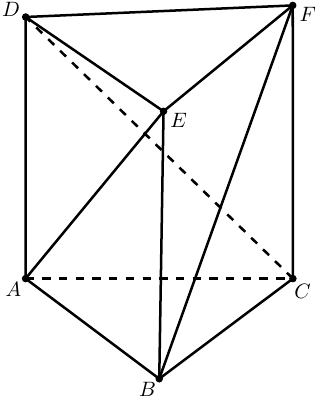}
\caption{The Sch\"{o}nhardt polyhedron $P$:
Almost a prism, except that the triangle $DEF$ is slightly rotated compared to $ABC$, forming concave edges $AE$, $BF$ and $CD$.
No tetrahedron spanned by four of the corners is contained in $P$, so $P$ cannot be tetrahedralized without adding a Steiner point.}
\label{fig:shonhardt}
\end{figure}

A \emph{non-convex} polyhedron may not have a tetrahedralization without adding Steiner points.
A well-known example where a Steiner point is necessary is the \emph{Sch\"{o}nhardt polyhedron}~\cite{schonhardt1928zerlegung}; see \Cref{fig:shonhardt}.
It is known that $\Theta(n^2)$ Steiner points and tetrahedra are sufficient and sometimes necessary~\cite{DBLP:journals/siamcomp/Chazelle84}.
Deciding whether or not Steiner points are needed is NP-hard~\cite{DBLP:journals/jal/BelowLR04}.

In order to obtain a compact volumetric model of a polyhedron, we would like to minimize the number of tetrahedra.
Somewhat surprisingly, adding a Steiner point may allow for fewer tetrahedra even when tetrahedralizing a convex polyhedron~\cite{DBLP:journals/dcg/BelowLR00}.
Finding the minimum number of tetrahedra without using Steiner points is NP-hard for convex polyhedra~\cite{DBLP:conf/soda/BelowLR00}.
Chin, Fung, and Wang \cite{DBLP:journals/dcg/ChinFW01} found a $2-\Omega(1/\sqrt n)$-approximation algorithm for this problem, which is optimal unless P$=$NP.
However, no algorithm is known of non-convex polyhedra.
This leads to our first question, which is also raised in~\cite{BernEppstein1995,DBLP:reference/cg/Bern04}.

\begin{quote}
\textbf{Problem~\CompactTetrahedralization}:
Given a polyhedron with $n$ vertices, can a tetrahedralization with Steiner points while using at most $O(\opt)$ tetrahedra be found time polynomial in $n$?
Here $\opt$ is the minimum number of tetrahedra in any tetrahedralization.
\end{quote}

In \emph{collision detection}, it is crucial to have a compact representation of the involved objects.
A common method to check whether three-dimensional objects intersect is to use \emph{bounding volume hierarchies}~\cite{DBLP:journals/tvcg/KlosowskiHMSZ98,CDChpaterLinManocha2018}.
A bounding volume hierarchy of an object $P$ is a rooted tree $T$ where each node $u$ corresponds to a subset $u(P)$ of $P$.
Here we will consider binary trees only, but trees of higher degrees are also often used~\cite{DBLP:journals/tvcg/SchmidtkeE18}.
The root $r$ of $T$ corresponds to all of $P$, i.e., $r(P)=P$.
For a vertex $u$ of $T$ with children $v_1,v_2$, the subsets $v_1(P),v_2(P)$ is a partition of $u(P)$.
The leaves of $T$ correspond to the geometric primitives of $P$, such as the triangles or tetrahedra defining $P$ using surface or volumetric models, respectively.
Several heuristics have been proposed for generating bounding volume hierarchies, including grouping the primitives according to when they are hit by a Morton space-filling curve~\cite{DBLP:journals/tvcg/SchmidtkeE18} or splitting the nodes recursively according to some heuristic~\cite{DBLP:conf/siggraph/GottschalkLM96,DBLP:journals/tvcg/KlosowskiHMSZ98}.
However, to the best of our knowledge, no methods have been described that find a hierarchy that is provably (close to) optimal according to any natural measure of optimality.
It is not clear what objective we should aim to optimize, but a good candidate is the total volume of the axis-aligned bounding boxes of all nodes.
The experiments of~\cite{DBLP:journals/tvcg/KlosowskiHMSZ98} suggest that this measure gives fast collision detection queries in practice:
In the paper, the splitting heuristic leading to the fastest queries is the one where we minimize the sum of the volumes of the two children.
This results in a hierarchy that is \emph{locally} optimal, but there may be other hierarchies with a smaller total sum over all nodes.
This leads to the next problem.

\begin{quote}
\textbf{Problem~\BoundingVolumeHierarchy}:
Give a polyhedron $P$ and a tetrahedralization $T=T(P)$, can an optimal bounding volume hierarchy, whose leaves are the tetrahedra of $T$,  be found in polynomial time, minimizing the sum of the bounding box volumes over all nodes?
\end{quote}

\begin{figure}
\centering
\includegraphics[page=11]{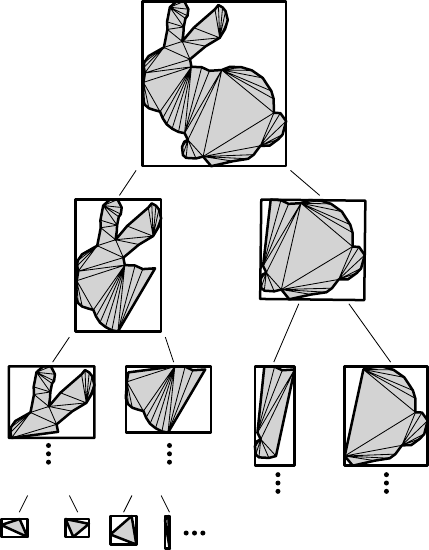}
\caption{Part of a bounding volume hierarchy of a triangulation of the silhouette of the Stanford Bunny.
The two-dimensional variant of problem~\BoundingVolumeHierarchy is to minimize the total area of the bounding boxes.}
\label{fig:optimalhierarchy}
\end{figure}

The problem is also interesting and apparently unexplored in the two-dimensional case; see \Cref{fig:optimalhierarchy} for an example.

\section{Partitioning large objects for manufacturing}

Consider a manufacturing process where we want to produce a large object $P$, but we are only able to produce parts of bounded size.
We then want to partition $P$ into parts that are sufficiently small to be fabricated, and in the end these can be assembled to form the desired object $P$.
To keep both fabrication and assembly as simple as possible, we naturally want as few parts as possible.
This problem for instance appears in 3D printing, where we can only print parts that fit in the printing volume, which typically has the shape of a box.
(For simplicity, we don't care whether the parts can in fact be assembled to form $P$ without colliding---considerations of the type discussed above in \Cref{sec:interlocked,sec:MDM}.)
Heuristics have been proposed to solve this problem~\cite{chen2022skeleton,jiang2017models,DBLP:journals/tog/LuoBRM12}, and several videos can be found on YouTube showing how it can be done manually using various software (for instance, search for \href{https://www.youtube.com/results?search_query=splitting+large+part+for+3D+printing}{``splitting large part for 3D printing''}), but no algorithm with provable guarantees on the number of resulting parts is known.

Abrahamsen and Stade~\cite{DBLP:journals/corr/abs-2404-09835} studied a two-dimensional version of this problem where the input is a simple polygon $P$ with $n$ corners.
They proved that it is NP-hard to partition $P$ into a minimum number of pieces each of which is contained in an axis-parallel unit square.
Abrahamsen and Rasmussen~\cite{DBLP:journals/corr/abs-2211-01359} described an approximation algorithm for this problem.
The algorithm runs in $O(n^2+\opt\log n)$ time and partitions $P$ into at most $13\,\opt$ pieces, where $\opt$ is the minimum number of pieces in any partition.
Abrahamsen and Rasmussen~\cite{DBLP:journals/corr/abs-2211-01359} also described $O(1)$-approximation algorithms for several related problems, such as partitioning into pieces that are contained in unit discs or have bounded diameter or perimeter.
The technique from~\cite{DBLP:journals/corr/abs-2211-01359} does not work for polygons with holes, and it seems even more difficult to devise an algorithm for polyhedra in $\mathbb R^3$, which is arguably the most important case from a practical perspective.
This leads us to the next problem.

\begin{quote}
\textbf{Problem~\PartitioningThreeD}:
Given a polyhedron $P$ in $\mathbb R^3$ with $n$ corners, is there an algorithm with running time polynomial in $n$ and $\opt$ for partitioning $P$ into $O(\opt)$ pieces, each of which is contained in an axis-parallel unit cube?
Here, $\opt$ denotes the minimum number of pieces in any such partition.
\end{quote}

\begin{figure}
\centering
\includegraphics[page=10]{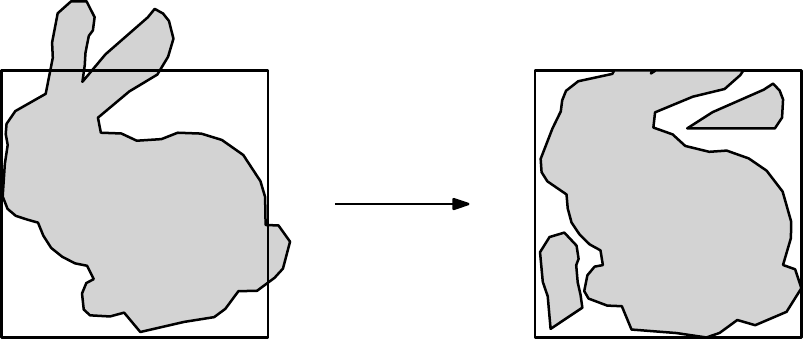}
\caption{The silhouette of the Stanford Bunny can be partitioned into three pieces that can be packed in the unit square after suitable rotations.}
\label{fig:partitioningforpacking}
\end{figure}

In 3D printing, it is often advantageous to print several parts simultaneously, if the parts can be packed into the printing volume.
Compared to printing the parts individually, in central industrial 3D printing methodologies, doing many at once is much more economical~\cite{YANG2023NESTINGIN3DPRINTING}.
Let us again consider a two-dimension setting where we wish to ``print'' a polygon $P$, and we can produce any set of polygons that can be packed into the unit square.
If $P$ has area exactly $1$, then the Wallace-Bolyai-Gerwien theorem\footnote{
See, e.g., the Wikipedia page ``Wallace-Bolyai-Gerwien theorem.''
}
says that it is possible to partition $P$ into finitely many triangles that can be packed into the unit square.
Here, the triangles are allowed to be rotated.
It follows that if the area of $P$ is at most $1$, it is always possible to partition $P$ into finitely many polygons that can be packed into the unit square.
Again, we naturally want to minimize the number of pieces, and we arrive at the following problem; see \Cref{fig:partitioningforpacking} for an example.

\begin{quote}
\textbf{Problem~\PartitioningPacking}:
Given a polygon $P$ with $n$ corners and area at most $1$, is there an algorithm with running time polynomial in $n$ and $\opt$ for partitioning $P$ into $O(\opt)$ pieces, so that these pieces can be simultaneously packed into the unit square?
Here, $\opt$ denotes the minimum number of pieces in any such partition.
We can allow the pieces to be rotated arbitrarily or allow translations only.
\end{quote}

Abrahamsen, Miltzow and Seiferth~\cite{DBLP:journals/corr/abs-2004-07558} proved that many packing problems are $\exists\mathbb R$-complete and thus likely not in NP.
This for instance holds for the problem where we are given a set of convex polygons and want to know if they fit in a unit square.
We expect that the problem~\PartitioningPacking is also $\exists\mathbb R$-hard to solve optimally for similar reasons, but will not attempt to prove it here.

Max Dehn proved in 1910 that the Wallace-Bolyai-Gerwien theorem does not hold in three dimensions~\cite{Zeeman_2002}:
He proved that the regular tetrahedron with volume $1$ cannot be partitioned into finitely many tetrahedra that fit in the unit cube.
However, if a polyhedron $P$ has volume at most $1-\varepsilon$ for a constant $\varepsilon>0$, then it is not hard to see that $P$ can be partitioned into finitely many pieces that can be packed into the unit cube:
We can partition $P$ using a regular grid of voxels of size $\delta\times\delta\times\delta$.
This results in some \emph{complete} pieces of $P$ that are cubes of size $\delta\times\delta\times\delta$ and some \emph{incomplete} pieces that are only subsets of such cubes.
Let us observe that there are at most $O(1/\delta^2)$ incomplete pieces:
Each facet of $P$ intersects $O(1/\delta^2)$ voxels and each of these intersections corresponds to a facet of an incomplete piece.
As there are $O(1)$ facets of $P$, it follows that there are $O(1/\delta^2)$ incomplete pieces.
Choosing $\delta=1/k$ for a sufficiently large integer $k$, all the complete pieces fit in a box of size $1\times 1\times (1-\varepsilon)$ in the bottom of the unit cube, since the volume of $P$ is also $1-\varepsilon$.
The remaining available space of the unit cube is a box of size $1\times 1\times \varepsilon$ with volume $\varepsilon=\Omega(1)$, so it can accommodate $\Omega(1/\delta^3)$ incomplete pieces and there is room for our $O(1/\delta^2)$ incomplete pieces when $\delta$ is small enough. 
We can thus also consider Problem~\PartitioningPacking in this three-dimensional setting with a bit of extra space in the unit cube, but it seems very difficult.

\section{Convex covering of free configuration space}

In many path planning scenarios, the free portion $P$ of the configuration space is difficult to handle directly.
Describing $P$ as a union of convex sets has shown to be a versatile technique, simplifying many such problems.
That is, we want to find a family $\mathcal C$ of convex sets such that\footnote{In general we may need infinitely many convex sets to cover $P$; we further refer to this issue below.} $P=\bigcup_{C\in\mathcal C} C$.
As a simple example, consider the intersection graph $G$ induced by $\mathcal C$, namely, the vertices of $G$ are the sets $\mathcal C$ and there is an edge in $G$ between two sets $C_1$ and $C_2$ if $C_1\cap C_2\neq\emptyset$.
In order to find a path between two points $p_1,p_2\in P$, we first find sets $u_1,u_2\in\mathcal C$ where $p_1\in u_1$ and $p_2\in u_2$.
We then find a path $\pi=v_1v_2\cdots v_k$ in $G$, where $v_1=u_1$ and $v_k=u_2$.
Letting $q_i\in u_i\cap u_{i+1}$, we then get a path $p_1q_1q_2\cdots q_{k-1}p_2$ from $p_1$ to $p_2$ in $P$, using the convexity of our sets.

Ideally, we want $\mathcal C$ to contain as few sets as possible, in which case finding $\mathcal C$ is known as the problem \emph{Minimum Convex Cover}.
This problem has a long history in Computational Geometry and is motivated by several other practical settings as well.
For instance, Garey and Johnson~\cite[p.~232]{garey1979computers} mentioned a very restricted version of the same problem: describing an orthogonal polygon as a union of a minimum number of axis-parallel rectangles.
They state that the problem is NP-hard and call the problem \emph{Rectilinear Picture Compression}, because such a set of rectangles is often a compact way to represent a set of pixels in an image.

\begin{figure}
\centering
\includegraphics[page=5]{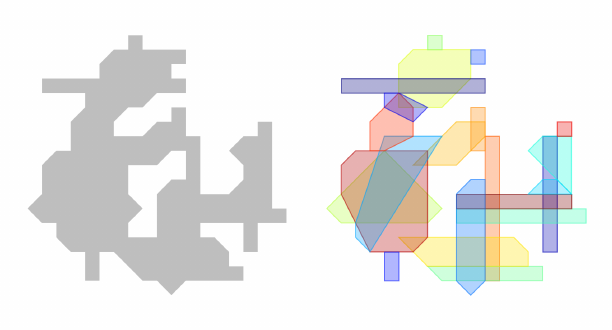}
\caption{A polygon (left) and a convex cover of the polygon (right), found with the method from~\cite{DBLP:conf/compgeom/AbrahamsenMN23}.}
\label{fig:CC}
\end{figure}

Marcucci \etal~\cite{DBLP:journals/scirobotics/MarcucciPWT23,DBLP:journals/siamjo/MarcucciUPT24} demonstrated that convex covers of the configuration space can be used to accomplish complex tasks in motion planning.
Motivated by this work, a two-dimensional version, where $P$ is a polygon with holes, was chosen for the 2023 edition of the challenge \emph{Computational Geometry: Solving Hard Optimization Problems}~\cite{DBLP:journals/corr/abs-2303-07007}.
The challenge was won by Abrahamsen, Meyling and Nusser~\cite{DBLP:conf/compgeom/AbrahamsenMN23}; see \Cref{fig:CC}.
They took a three-step approach, where they first triangulated $P$ appropriately.
Then they computed a visibility graph on the triangulation, where there is an edge between two triangles when any point in one triangle can see any point in the other.
Finally, a small clique cover of the visibility graph was found, and each clique was turned into a convex piece of the resulting cover.
Inspired by this, Werner \etal~\cite{DBLP:journals/corr/abs-2310-02875} used a similar approach for configuration spaces $P$ of higher dimension.
Instead of triangulating $P$ (which may not be explicitly known in more complicated path finding settings), they sampled sufficiently many points from $P$.
They then computed the visibility graph of the sampled points, found a clique cover, and then grew ellipsoids around each clique.

We will formulate a question about the two-dimensional version of the problem, that is, $P$ is a connected region in the plane, possibly with holes.
However, the question is also relevant in higher dimensions.
Likewise, versions of our problem can be formulated when the configuration space is not known explicitly, but we can sample points from it.
It is very difficult to find minimum convex covers:
the problem of writing a simple polygon as a union of a minimum number of convex polygons has even been shown to be $\exists \mathbb R$-complete~\cite{DBLP:conf/focs/Abrahamsen21}, and it is thus likely not in NP.
A constant-factor approximation algorithm is known for this case~\cite{DBLP:conf/focs/BrowneKMP23}.
In practice, it is often not feasible or necessary to cover all of $P$.
For instance, if a concave curved arc appears on the boundary of $P$, then infinitely many convex sets are needed to cover $P$.
However, we will insist that $\bigcup_{C\in\mathcal C} C$ is connected and is sufficiently similar to $P$.
As a measure of similarity, we will here use the \emph{Hausdorff distance} $d_H$.
For two sets of points $A,B$, we define
\[
d_H(A,B)=\max\{\sup_{a\in A} \newinf_{b\in B} \lVert ab\rVert, \sup_{b\in B} \newinf_{a\in A} \lVert ab\rVert\},
\]
i.e., $d_H(A,B)$ is the largest distance from a point in $A$ or $B$ to the other set.
We can now formulate our first problem related to convex covers.

\begin{quote}
\textbf{Problem~\ConvexCover}:
Given a tolerance $\varepsilon>0$ and a polygon $P$ with $n$ vertices (and possibly with holes), devise an algorithm to find a family $\mathcal C$ of convex sets with the following properties, where $D=\bigcup_{C\in\mathcal C} C$:
\begin{itemize}
\item $D$ is connected,
\item $D\subseteq P$, and
\item $d_H(D,P)\leq \varepsilon$.
\end{itemize}
The running time should be polynomial in $n$ and $1/\varepsilon$.
We want to minimize the number of sets $\vert \mathcal C\vert$; can we for instance find a set $\mathcal C$ that is only a constant factor larger than the optimal such set?
\end{quote}

\begin{figure}
\centering
\includegraphics[page=7]{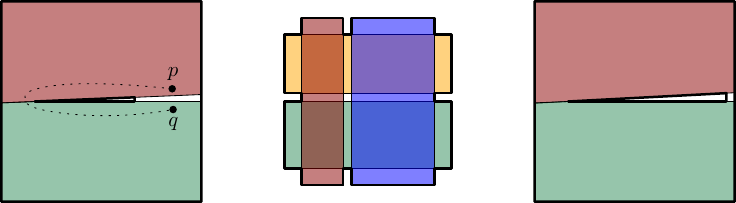}
\caption{Left: Here, $P$ is a square with a thin triangular hole.
When using the standard Hausdorff distance $d_H$ as similarity measure, the optimal solution results in a long uncovered corridor, leading to a long detour between the points $p$ and $q$.
Middle: A small hole in the cover as shown here may be acceptable as it does not significantly impair the connectivity of the region when using the cover for path planning tasks, but the stronger variant $d_{H^*}$ will prevent that, which may result in more convex sets than are actually needed.
Right: Even with $d_{H^*}$, narrow passages in $P$ may be lost in the convex cover, but this is often acceptable in practice since a path using a narrow passage will have small clearance, which is undesirable (see also the last problem of \Cref{sec:shortest-coordinated}: ~\TwoDiscsMinCombinedMeasure).
}
\label{fig:Hausdorff}
\end{figure}

One might consider other similarity measures as well, instead of just the ordinary Hausdorff distance.
An interesting variation is to consider
\[
d_{H^*}(A,B)=\max\{d_H(A,B),d_H(A^c,B^c)\},
\]
where we also consider the Hausdorff distance between the complements.
The optimal solution to the problem \ConvexCover with $d_H$ sometimes changes the topology of $P$ in an undesirable way, giving up too much on connectivity, that does not happen when using $d_{H^*}$.
On the other hand, $d_{H^*}$ also has the downside of sometimes being too strict, thus requiring an excessive number of sets in the cover; see \Cref{fig:Hausdorff}.

As another version of the problem, we can also aim to minimize the total number of vertices of the sets $\mathcal C$, instead of the number of sets.
This can be seen as the problem of finding a simpler representation of $P$.

\printbibliography

%\bibliographystyle{abbrv}
%\bibliography{ten}

\end{document}